\documentclass[aps,preprint,pre,array,epsfig,eqsecnum]{revtex4}

\usepackage{amsmath}
\usepackage{amsfonts}
\usepackage[dvips]{graphicx}
\usepackage{epsfig}
\usepackage[]{subfigure}
\usepackage[dvips]{color}

\begin{document}   
\setlength{\parindent}{0pt}

\title{Competing Polymerization of Actin Skeleton explains Relation between
Network Polarity and Cell Movements}

\author{B. Nandy and A. Baumgaertner}

\affiliation{
Institut f\"ur Festk\"orperforschung, 
Forschungszentrum J\"ulich, 52425 J\"ulich, Germany}

\date{\today}

\begin{abstract} 

Based on experimental observations it is known that various biological
cells exhibit a persistent random walk during migration on flat substrates.
The persistent random walk is characterized by `stop-and-go' movements :
unidirectional motions over distances of the
order of several cell diameter are separated by localized short time
erratic movements.
Using computer simulations the reasons for this phenomena had been unveiled and
shown to be attributed to two antagonistic nucleation processes during
the polymerization of the cell's actin cytoskeleton : the (ordinary) spontaneous
nucleation and the dendritic nucleation processes.
Whereas spontaneous nucleations generate actin filaments growing in different 
directions and hence create motions in random directions, 
dendritic nucleations provide a unidirectional growth.
Since dendritic growth exhibits stochastic fluctuations,
spontaneous nucleation may eventually compete or even dominate, 
which results in a reorientation of filament growth and hence 
a new direction of cell motion.
The event of reorientation takes place at instants of vanishing
polarity of the actin skeleton.

\end{abstract}


\maketitle

\section{INTRODUCTION}

Most animal cell types are `motile' and 
possess the capacity to move over or through 
a substrate, and cell migration is a normal occurrence in both normal
physiological cases as in the case of diseases.
Unlike the phenomena related to `mobility', where the collision among molecules
determine their random movements, the random motion of motile cells
requires an energy-consuming mechanism. 
It is now widely accepted that the basic engine for gliding or 
crawling locomotion of many living cells is the ATP-supported 
polymerization of the actin cytoskeleton.
Rapidly moving cells 
can often move over the substrate without microtubules (keratocytes,
neutrophils), but crawling motility always requires actin. 

One of the crucial 
factors for cell movement is actin polarity \cite{Weiner02}, 
which correlates with a 
persistent random walk during migration.
The persistent random walk is characterized by `stop-and-go' movements :
unidirectional motions over distances of the
order of several cell diameters are separated by localized short time
erratic movements.
The main focus of the present study 
is the cause behind the persistent random walk of a cell, and hence the
cause behind the correlated spontaneous change of polarity.

It has been seen that cell motility and chemotactic 
migration are related to specific rearrangements of the actin cytoskeleton.
Actin 
polymerizes into new filaments in regions of the cell that are specified by 
activated signalling at the plasma membrane
\cite{Bray92,Lauffenburger96,PollardBorisy03,Carlier03}. These signals give
rise to cell polarity and directional motility. In the absence
of chemotactic stimuli, cells exhibit a persistent random walk.

In this work we present results of our investigations on the
relation between cytoskeletal actin polymerization and the 
persistence random walk of a cell.
The results are based on the analysis of
Monte Carlo simulations of a simple model cell. 

\section{MODEL AND SIMULATION TECHNIQUE}

Due to the limitation of information and computational complexity it 
is not possible to construct the model cell and it's motility including all 
regulatory proteins with their exact concentration for all 
types of cells. So we focus on one of the simplest cell which is the 
keratocyte residing in the epidermis, the most outer layer of the skin.
The keratocyte is responsible for the formation of tissue and wound 
healing, both requiring the cell's ability to autonomously migrate 
within skin tissue.
It is in fact one of the fastest moving cells \cite{Alberts98}
with a speed of 0.5 $\mu$m/s. The advantage of choosing keratocyte is that
the crescent-like shape of the cell 
is almost constant as it moves \cite{Mitchison91,Lee93}.
The motility of the keratocyte is assumed to be based largely on a
continuously remodeling actin network. 
In our model we neglect the cell body, i.e., the
nucleus and other organelles. This cell type is known as a cytoplast
\cite{Borisy99a}.

\noindent
\subsection{The description of Membrane and Actin Models}

{\bf Membrane.}~~
The plasma membrane of a biological cell is a highly complex surface 
consisting of a lipid bilayer. The complexity of the cell membrane 
cannot be captured in our simple model membrane. Therefore, similar to the 
previous successful studies \cite{Leibler89,Sambeth01b,Satya04}, the 
cell membrane is designed as an elastic two-dimensional ring.
Our model membrane is a flexible 
closed ring embedded on the square lattice. The ring is 
non-self-avoiding
and exibit the usual random walk characteristics.
Conformational changes 
of the ring are achieved by Monte Carlo methods, where a randomly chosen pair 
of two successive segments of the chain perform a kink 
jump or hair pin jump to one of the neighbouring sites.

{\bf Globular Actin (G-actin).}~
The model membrane encloses a fixed number, $N$, of actin molecules. 
Each actin molecule has a size \cite{Alberts98}
of about 5 nm $\times$ 5 nm
and is located at any of the vertices of the square lattice. 
The G-actin molecules  diffuse freely from one lattice point to another. 
No excluded volume condition is imposed among G-actin molecules.
But it is imposed between membrane and G-actins. So the membrane is 
impermeable for all actin molecules.

{\bf Filament Actin (F-actin).}~
The G-actin monomers form a rigid filament by associating with each 
other. Within a filament the actin  monomers are called F-actin.
According to experimental results \cite{Horwitz02,DeMali03a},
the filaments are assumed to be chemically coupled to the
underlying substrate via membrane proteins. 
This attachment to the substrate provides the necessary traction forces
for the advancement of the cell.
Hence, in our model the actin filaments are assumed to be immobile.
Excluded volume effects between filaments and diffusing G-actin molecule 
are neglected, whereas excluded volume effects among filaments and between
filaments and membrane are included.

{\bf Actin-associated Proteins.}~
A group of actin associated proteins help the remodelling
of the actin network, and hence control membrane protrusion.
Proteins of the WASP/Scar families activate Arp2/3 protein complexes.
These complexes nucleate new actin filaments at the sides of existing filaments.
The nascent filaments in the network arise in the form of 
branches from preexisting filaments. 
The newly formed filaments elongate at their ends, pushing the 
membrane at the leading edge forward until they are capped by specific proteins.
For example,  
ADF/Cofilin creates free ends by severing preexisting filaments and 
promoting depolymerization of free filaments 
at their opposite ends, and 
CapZ and gelsolin cap the fast growing  ends.

Based on experimental observations it is known that Arp2/3 
is activated only close to the membrane. Therefore Arp2/3-induced
branching processes must be expected to happen only near the membrane.
The Arp2/3 molecule is not explicitly taken into account in our model
because of its large physiological concentration, in particular near the
cell membrane where it becomes activated. In our model, the width of the range 
of activation is taken to be 10 lattice sites. 
Accordingly, in our model 
we perfom branching from an existing filament
only if the filament extends to the range of activation near the membrane. 

A typical snapshot of the simulated cell 
is depicted in Fig.1.

\begin{figure}
\begin{center}
 \subfigure[]{\includegraphics [width=0.70\textwidth,angle=0]
{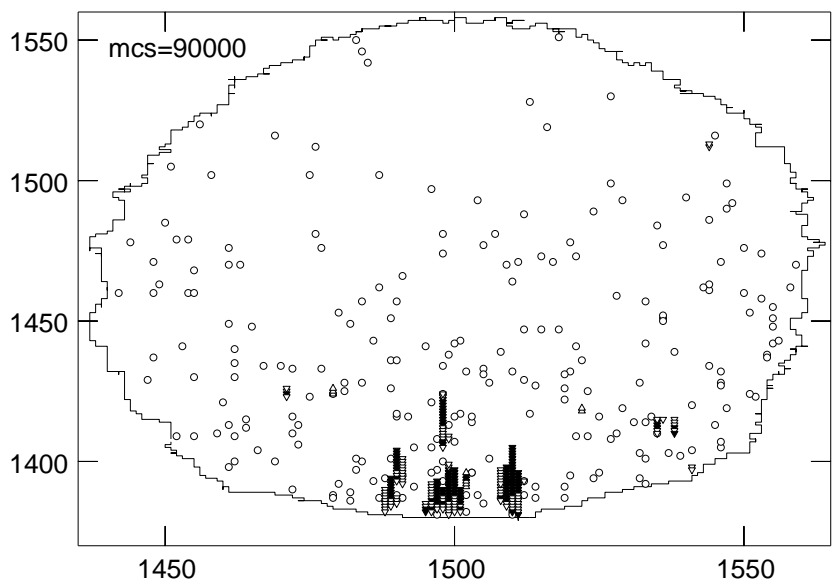}}\\
\subfigure[]{\includegraphics [width=0.70\textwidth,angle=0]
{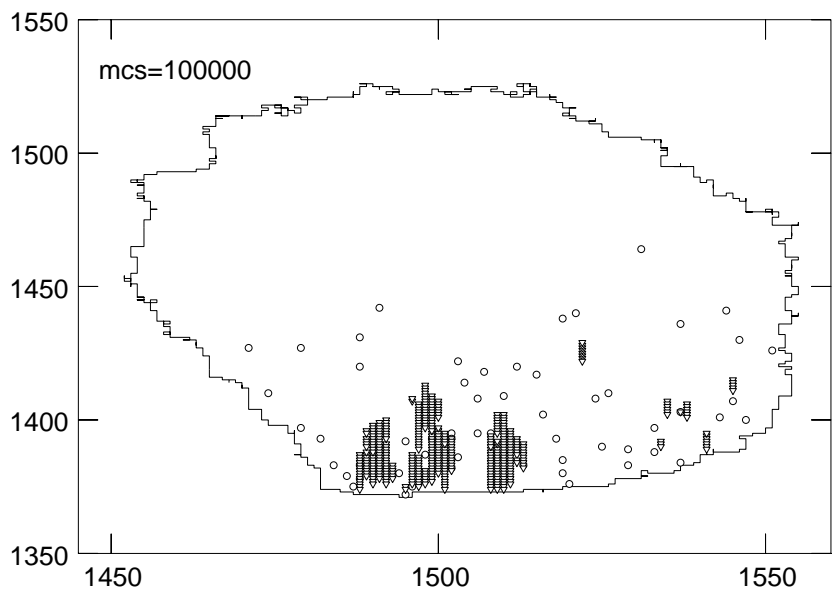}}
 \label{fig:snapshot}
 \caption{
 Two snapshot ($a$ and $b$) of the model cell. The open circles represent
    G-actin, the filled triangles represent F-actin molecules.     
    The length
    of the membrane is L=200, the total number of actin molecules is N=500.
   }
  \end{center}
  \end{figure}

\noindent
\subsection{Probabilities and Reaction Rates}

\vskip0.2cm\noindent
{\bf Polymerization.}~
During the polymerization step the association of G-actin molecules 
to an existing F-actin filament occurs at certain rates at both ends.
We adopt the same reaction rates for polymerization and depolymerization
as used in a previous work \cite{Satya04}, and are listed in Table I.
Due to the structural rotational symmetry of a single G-actin molecule,
both ends of the F-actin filament are distinguishable : the `barbed'
(or synonymously, `plus' or `fast growing') end, and the `pointed'
(or synonymously, `minus' or `slow growing') end.
The association rates of ATP-bound G-actin to both ends of a filament
are different \cite{Bray92,Alberts98,Wegner76,Wang85}, and we denote
the rate constants $k^+_b$ and
$k^+_p$ for `barbed' and `pointed' ends, respectively. 
The same holds for the
depolymerization rates $k^-_b$ and $k^-_p$, albeit their difference
is smaller.
Since both ends of a filament are distinguishable, 
each F-actin filament possesses an intrinsic polarity.

\begin{table}\label{tab:model}
{\footnotesize
\begin{tabular}{|l|c|}
\hline
    Quantity                &  Value         \\ \hline
  lattice constant $a$ &  5 nm                         \\
  typical cell size & 100 $\times$ 100 lattice \\
  Monte Carlo step $\tau$  &  0.875 $\mu$s  \\
  $W_n$                  & 0.01   \\
  $W_{br}^+$               & 0.1   \\
  $W_{br}^-$               & 0.6  \\
  $W_b^+$                & 1    \\
  $W_b^-$                & 0.0012     \\ 
  $W_p^+$                & 0.11   \\
  $W_p^-$                & 0.02    \\ \hline   
\end{tabular}
\caption{Model parameters and reaction probabilities.}  
}
\end{table}

\vskip0.2cm\noindent
{\bf Nucleation.}~
We consider two types of nucleation for F-actin.
When nucleation is formed by two G-actin molecules,  it is called `spontaneous' 
nucleation which occurs with probability $W_n$. When
the `branching' nucleation takes place then a new filament is formed
as a branch from the side of an existing filament.
The branching nucleation is modeled
following the `dendritic nucleaction' model based on experimental observations
\cite{Borisy99b,Pollard01,Pollard02}.
It is known 
that the activated protein complex Arp2/3 
can associate with an existing
filament and can  nucleate there a new filament
as a branch from the mother filament at an angle of about 70$^{\circ}$. 
This leads to the formation of a branched network. 
In our model on the square lattice, the branching process is implemented
as follows. 
If a G-actin is found to be on the adjacent row to an 
existing filament, a new daughter 
filament is created at this site with probability $W^{+}_{br}$. 
If there 
is already a filament in that row, then the nucleation attempt is rejected.
Based on the experimental facts \cite{Borisy99b,Pollard01,Pollard02}, the
tip of the daughter filament is a plus end, 
and hence its
direction of growth must be the same as the plus end of the
mother filament.
This implies, that the polarity of a filament, i.e., the vector connecting
plus and minus ends, is determined by the spontaneous nucleation
process and cannot be changed by dendritic nucleation.
In our model,
for the sake of simplicity, the spontaneous nucleation creates
new filaments only parallel to the Y-axis, pointing at random in
one of the two directions.

\vskip0.2cm\noindent
{\bf Reaction Rules.}~
Regarding the reaction rules we follow our earlier work \cite{Satya04}. In brief, 
the reaction rules are as follows. First we randomly choose one actin 
molecule. Then for the selected actin molecule, a random choice with 
a equal probability is made between an association or dissociation 
process. Then four cases may happen.
1) If  for a G-actin molecule the dissiocation process is selected, the step is 
stopped.
2) If the accociation process is selected for G-actin molecules, then reaction 
may happen provided the neighbour G or F-actin exist and the 
probability $W_n$ or $W^{+}_{br}$  or $W^{+}_p$ or $W^{+}_b$  fullfill the 
condition $W > \eta$, where $\eta$ is a random number $0 < \eta < 1$.
3) If the association process for an F-actin is chosen, then a successful 
reaction may happen provided a G-actin as a neighbor exists. 
4) If a dissociation process of a F-actin molecule is selected, then the 
process will take place with probability $W^{-}_b$ or $W^{-}_p$ or $W^{-}_{br}$.  .

\vskip0.2cm\noindent
{\bf Reaction Probabilities and Monte Carlo steps.}~
Since the experimental reaction rates $k^{\pm}_{b,p}$ cannot be used directly 
in simulations, 
but instead reaction probabilities $W^{\pm}_{b,p}$,
one must establish a relation between them.
In order to calculate these probabilities, 
we define $W^{+}_{b}$ = 1 for barbed end polymerization 
and calculate $W^{+}_{p}$
for the pointed end using the relation
\begin{equation}
\label{eq:Wk}
\frac{W^+_p}{W^+_b} = \frac{k^+_p}{k^+_b}~,
\end{equation}
The same can be done for the depolymerization rates $W^{-}_{b,p}$, which we
choose $W^-_p$=0.02 and calculate $W^-_b$ by using the corresponding relation
to Eq.(\ref{eq:Wk}).
The value of $W^-_b$ has been choosen in order to ensure that
the filaments are much shorter than the cell size during the simulations.
A basic question is how many Monte Carlo steps, $m$, for Brownian motion 
membrane and G-actins
have to be performed between two sucessive chemical rection attempts.
Based on the experimentally known reaction rates and the diffusion coefficient
of G-actins, a detailed analysis 
\cite{Satya04} have shown that $m$ =  5 is an appropriate value.
All model parameters are summarized in Table I.

\section{Results and Discussions}

\subsection{Persistent Random walk}

\begin{figure}[h] 
  \begin{center}
    \includegraphics[width=0.70\textwidth,angle=0]
{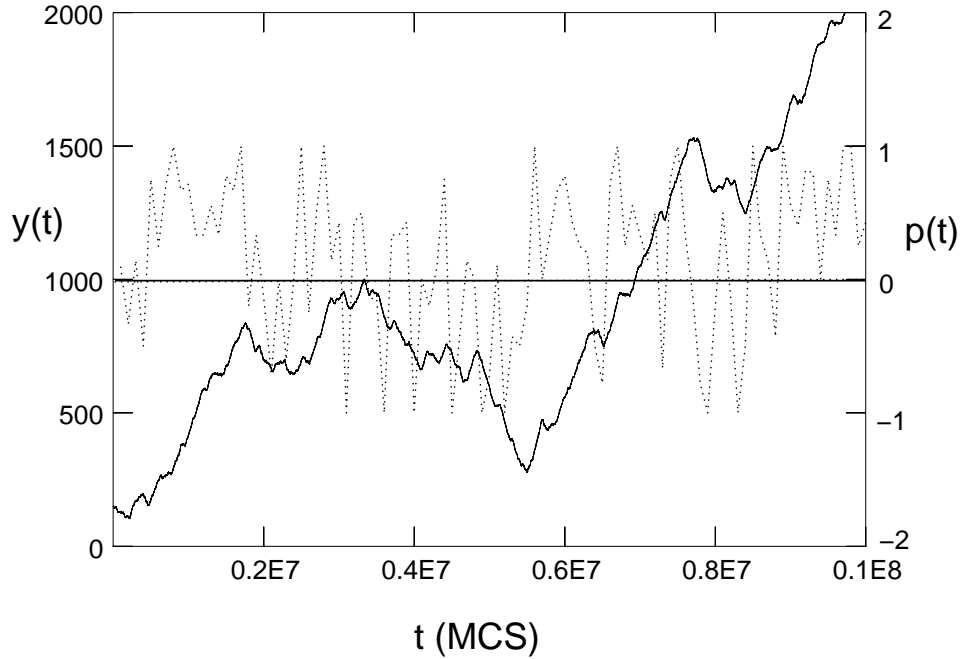}    
    \caption{The trajectory $Y(t)$ (full line, left scale) 
    and polarity $P(t)$ (dotted line, right scale)
    of a cell exhibiting a persistent random walk, as function
    of time $t$ in units of Monte Carlo steps, MCS. } 
    \label{fig:prw_d1}
  \end{center}
\end{figure}

\begin{figure}[h] 
  \begin{center}
    \includegraphics[width=0.75\textwidth, angle=0]
{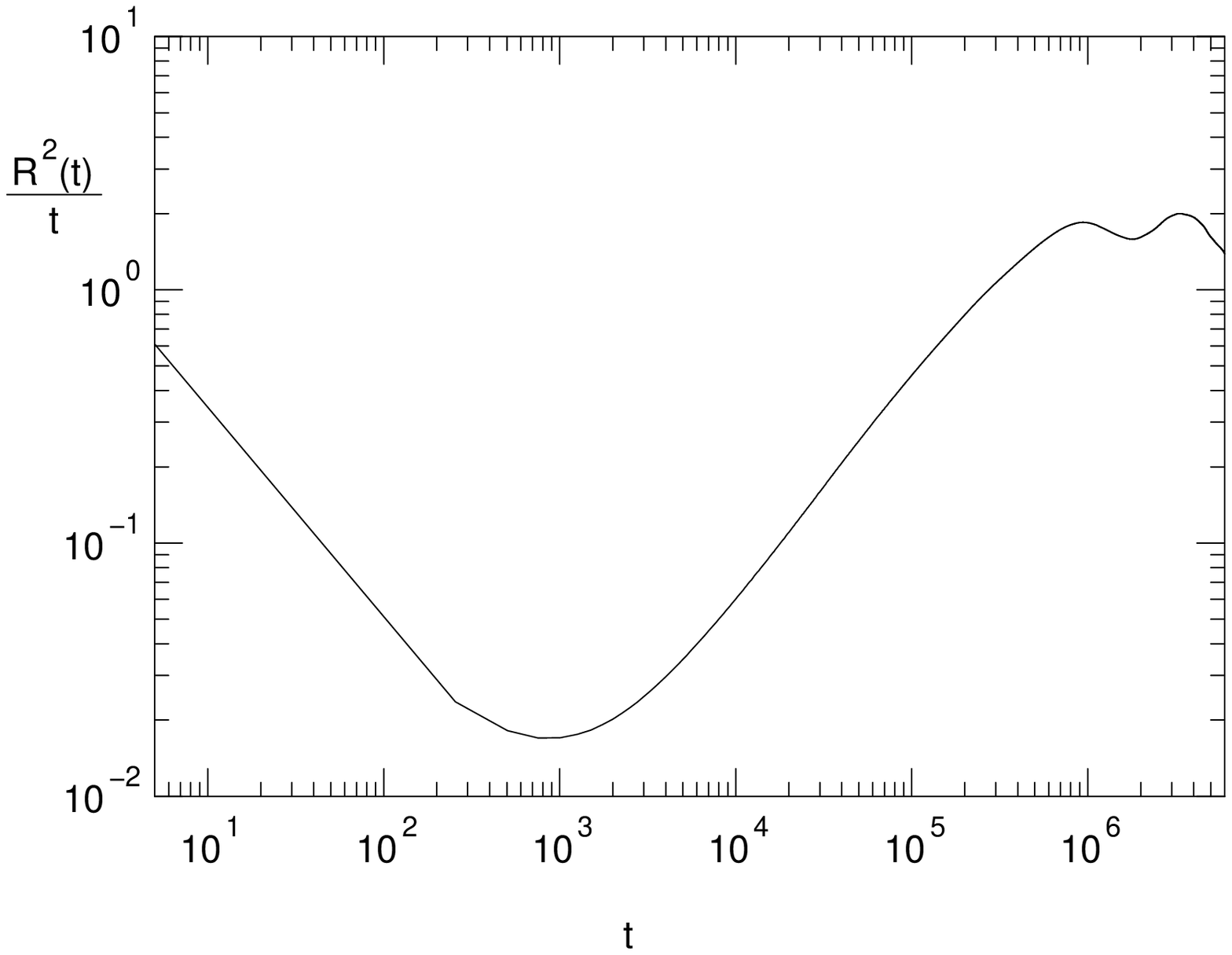}    
    \caption{Typical mean square displacement of a cell 
    exhibiting a persistent random walk.} 
    \label{fig:msd_d2}
  \end{center}
\end{figure}

From experimental observations \cite{Bray92,Lauffenburger96,Shreiber03} and
from simulation studies \cite{Sambeth01a,Satya04}, it is known that
Arp2/3-induced dendritic nucleation is essential for the appearance of
a persistence random walk (PRW) of certain cells. 
The PRW is characterized by `stop-and-go' movements :
unidirectional motions with almost constant velocity
over distances of the
order of several cell diameters are separated by localized short time
erratic movements. After each interruption of the ballistic motion, the
cell continues to move in a different direction. 
A typical trajectory of the center of our model cell is
shown in Fig.\ref{fig:prw_d1} as an example from our simulations.
The persistency of the cell's random walk can be deduced 
from the time-dependent mean square displacement of the cell 
\begin{equation}
     R^2(t) = \langle [Y(t) - Y(0)]^2 \rangle
\end{equation}
which is shown in Fig.\ref{fig:msd_d2}. 
The dynamics at short and long times are 
governed by diffusion, whereas for intermediate times, 
the cell motion exhibits a unidirectional drift during a certain
persistence time.
The unidirectional motion has been explained as a consequence of 
the autocatalytic nature \cite{Carlier00,Sambeth01a} of 
the dendritic polymerization kinetics. 
The branching processes induce
an explosive growth of filaments in a certain direction which 
leads to membrane protrusions and concomitant cell movements in the
same direction.
%
%
%
This motion lasts only for a certain persistence time, then
the direction of motion changes. 

In the following subsection we discuss 
the reasons for this dynamic instability.
It should be noted that insights to the physical basis of the cell's
persistency 
would support, in particular, our
understanding of the physical basis of chemotaxis, i.e., the prolongation
of the persistency of the cell's motion by an external signal.

\subsection{Dynamic Instability}

The fundamental quantity which governs the migration of the cell is its
polarity and its mechanism of self-perpetuation. 
The cell polarity is 
macroscopically and dynamically characterized by the
formation of leading and trailing membrane edges 
which are a consequence of the cell's coupling to the polarity of the 
enclosed actin cytoskeleton. 
The polarity of the cytoskeleton itself is the average orientation
of its filaments, where each actin filament possesses a polarity due to
its plus and minus ends.
This spontaneous internal polarity of the actin network
determines the direction of cell motion
and is maintained without external signals (e.g. chemotaxis) only for
a certain time (persistent random walk).
Specifically, the formation and regulation of cell polarity is achieved by
a complex protein signaling network with positive and negative feedback
loops. 
We do not discuss the details here which are 
beyond the scope of this paper. We note that different signals converge,
among others, on
the activation of an Arp2/3 protein complex, which leads to branching and 
the autocatalytic polymerization.

In our model, we define the polarity of the cell by the polarity of the total
actin network, i.e., 
by the difference 
in the number of filaments pointing in opposite directions,
\begin{equation}
     P = \frac{(F_{up}-F_{dn})}{F_{total}}     
\end{equation}
where $F_{up}$ is the number of filaments pointing in 
positive Y-direction, $F_{dn}$ in negative Y-direction, and
$F_{total}$ is the total number of filaments.
The range of the polarity parameter is  $-1\le P \le 1$.
The trajectory of this quantity is shown
in Fig. \ref{fig:prw_d1} (dotted line, right scale). 
From this result one observes  
that at certain times, where $P(t) = 0$, 
the cell changes its direction of motion to the opposite as
indicated by the corresponding displacements $Y(t)$. 

After the analysis of several simulations under
different conditions of the model cell, it must be concluded that the  
cause behind this phenomena is the existence of two antagonistic processes, 
the spontaneous and the dendritic nucleation processes.

During a spontaneous
nucleation process, where two G-actin monomers form a new
F-actin filament, the orientation of the new filament is determined at random.
This is in contrast to the dendritic
nucleation process, where the daughter
filament branches off the mother filament and therefore
always adopts the same growth direction as the mother filament.
Hence, spontaneous nucleation changes the sign of polarity, whereas
dendritic nucleation increases the absolute value of polarity.
The first leads to random motion, whereas the latter prolongs 
unidirectional motion.
Since the polarity fluctuates in time, the spontaneous nucleation
may eventually dominate and causes a change of polarity, and hence
a change of vectorial motion.

\begin{figure}[h] 
  \begin{center}
    \includegraphics[width=0.76\textwidth, angle=0]
{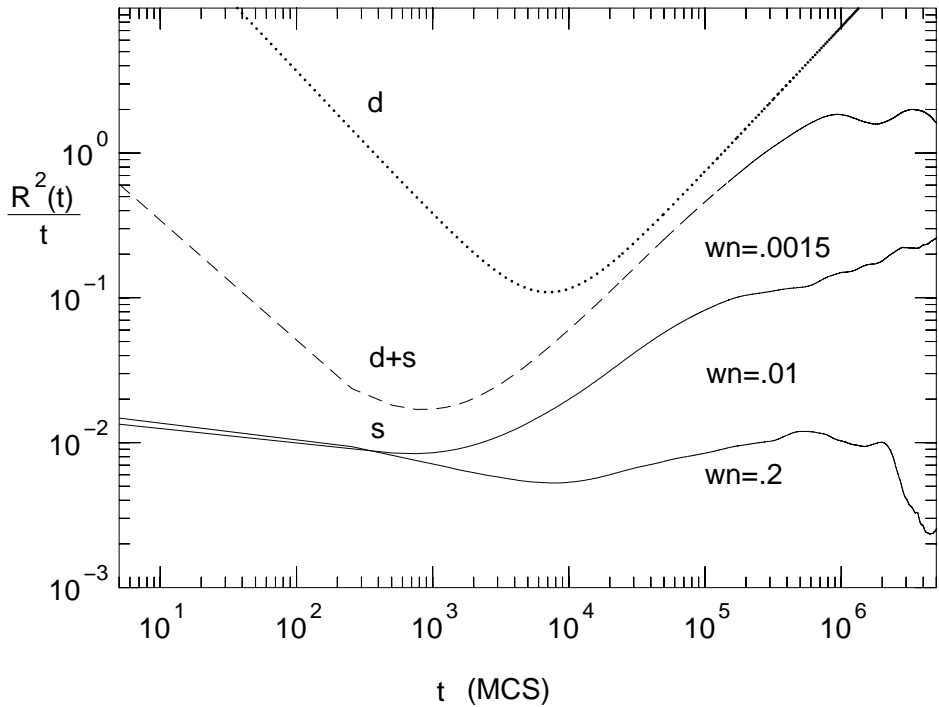}    
    \caption{Mean square displacements of a cell under different 
    conditions and at different probablilities of spontaneous nucleation, $wn$.
     `s' represents spontaneous nucleation, `s+d' represents both 
     spontaneous and dendritic nucleation, and `d' represents  dendritic
     nucleation only.}  
    \label{fig:msd}
  \end{center}
\end{figure}

This view is corroborated by simulations of various limiting cases, which are
shown in Fig.\ref{fig:msd}.
%

In one limiting case, the cell motion is caused solely by spontaneous
nucleations.
The corresponding curve 
is shown in Fig.\ref{fig:msd} by the full lines with label `s'.
At two different nucleation probabilities, $W_n$, the mean square
displacements indicate
ordinary diffusion, $R^2(t)/t \approx const.$, at almost all times.
This is expected due to the random orientation of newly established
filaments. A very weak persistency may be expected at small nucleation rates.

In the other limiting case, the cell motion is caused solely by dendritic 
nucleation. Since in this case the orientation of the filaments is preserved,
the corresponding curve (dotted line with label `d') has to
exhibit a clear drift, $R^2(t)/t \sim t$, which takes place  for
times larger than a characteristic time $\tau_d \approx$ 7 $\times$ 10$^3$. 
The characteristic time scale $\tau_d$ separates periodes of cell 
advancements and localized cell displacements (`resting').
Since the activation range of Arp2/3 extends only a few lattice sites from
the membrane, the rapid advancement of the leading edge of the membrane  
together with the rapid autocatalytic processes of branching near
the leading edge deprive this area of available G-actin molecules which are
necessary for continuation of these processes.
Therefore, actin network and cell remain temporarily, during $t < \tau_d$,
essentially at rest.
While recruiting G-actins by diffusion
from the minus to the plus ends of the filaments near the activation zone, 
the membrane of the cell performes random displacements 
around the immobile filament network, which leads to time-independent
displacements of the cell, $R^2(t)/t \sim 1/t$ at $t < \tau_d$. 
It is still unknown whether this result from simulations
is related to the experimentally observed \cite{Vicker00}
excitation waves of F-actin assembly near the membrane and the correlated
cell advancements. 

The interesting intermediate case, when a few events of spontaneous nucleation 
interfere with dendritic nucleations, is shown in Fig.\ref{fig:msd}
by the broken line with label `d+s'.
At a nucleation ratio of $W_n/W_{br}^+ = $0.015, 
again a drift of the cell is observed albeit only up to
a typical persistence time $\tau_p \approx 10^6$.
At larger times, ordinary diffusion $R^2(t)/t = const.$ is observed
indicating the dominance of spontaneous nucleation processes.
For $t < \tau_p$, the mean square displacements are qualitative the same as
for the dendritic case `d'.

\section{Summary and Conclusions}

The dynamics of a motile cell with unidirectional and bidirectional branching
is studied with treadmilling actin polymerization kinetics. The filaments 
growing on both sides with treadmilling rates and branching in both directions
leads to a persistent random walk whose 
persistency becomes infinite and leads to unidirectional drift
in the case of one side branching.
The dynamics are governed by  
diffusion at short times followed by drift at longer times. 

The phenomena is 
shown to be attributed to two antagonistic nucleation processes during
the polymerization of the cell's actin cytoskeleton : the (ordinary) spontaneous
nucleation  and the dendritic nucleation  processes.
Whereas spontaneous nucleations generate actin filaments growing in different 
directions and hence create motions in random directions, 
dendritic nucleations provides a unidirectional growth.
Since dendritic growth exhibits stochastic fluctuations,
spontaneous nucleations may eventually compete or even dominate, 
which results in a reorientation of filament growth and hence 
a new direction of cell motion.
The event of reorientation takes place at instants of vanishing
polarity of the actin skeleton.

\section*{ACKNOWLEDGEMENTS}
B. Nandy would like to thank IBM Research for providing a doctoral
fellowship and
the Institut f\"ur Festk\"orperforschung
at the Forschungzentrum J\"ulich for hospitality.


\bibliographystyle{phaip}
\bibliography{Nandy}

\begin{thebibliography}{10}

\bibitem{Weiner02}
D.~Weiner,
\newblock Curr. Opin. Cell Biol. {\bf 14}, 196 (2002).

\bibitem{Bray92}
D.~Bray,
\newblock {\em Cell Movements},
\newblock Garland Publ., New York, 2nd. edition, 2001.

\bibitem{Lauffenburger96}
D.~A. Lauffenburger and A.~F. Horwitz,
\newblock Cell {\bf 84}, 359 (1996).

\bibitem{PollardBorisy03}
T.~D. Pollard and G.~G. Borisy,
\newblock Cell {\bf 112}, 453 (2003).

\bibitem{Carlier03}
M.-F. Carlier, C.~L. Clainche, S.~Wiesner, and D.~Pantaloni,
\newblock BioEssay {\bf 25}, 336 (2003).

\bibitem{Alberts98}
B.~Alberts et~al.,
\newblock {\em Essential Cell Biology},
\newblock Garland Publishing, Inc., New York, 1998.

\bibitem{Mitchison91}
J.~Theriot and T.~Mitchison,
\newblock Nature {\bf 352}, 126 (1991).

\bibitem{Lee93}
J.~Lee, A.~Ishihara, J.~Theriot, and K.~Jacobson,
\newblock Nature {\bf 362}, 167 (1993).

\bibitem{Borisy99a}
A.~B. Verkhovsky, T.~M. Svitkina, and G.~G. Borisy,
\newblock Curr. Biol. {\bf 9}, 11 (1999).

\bibitem{Leibler89}
S.~Leibler, R.~P. Singh, and M.~E. Fisher,
\newblock Phys. Rev. Lett. {\bf 59}, 1989 (1989).

\bibitem{Sambeth01b}
R.~Sambeth and A.~Baumgaertner,
\newblock J. Biol. Systems {\bf 9}, 201 (2001).

\bibitem{Satya04}
S.~Satyanarayana and A.~Baumgaertner,
\newblock J.Chem.Phys. {\bf 121}, 4255 (2004).

\bibitem{Horwitz02}
D.~J. Webb, J.~T. Parsons, and A.~F. Horwitz,
\newblock Nature Cell Biol. {\bf 4}, 97 (2002).

\bibitem{DeMali03a}
K.~A. DeMali and K.~Burridge,
\newblock J. Cell Sci. {\bf 116}, 2389 (2003).

\bibitem{Wegner76}
A.~Wegner,
\newblock J. Mol. Biol. {\bf 108}, 139 (1976).

\bibitem{Wang85}
Y.~L. Wang,
\newblock J. Cell Biol. {\bf 101}, 597 (1985).

\bibitem{Borisy99b}
T.~M. Svitkina and G.~G. Borisy,
\newblock J. Cell Biol. {\bf 145}, 1009 (1999).

\bibitem{Pollard01}
K.~J. Amann and T.~D. Pollard,
\newblock Nature Cell Biol. {\bf 3}, 306 (2001).

\bibitem{Pollard02}
T.~D. Pollard and C.~C. Beltzner,
\newblock Curr. Opin. Struct. Biol. {\bf 12}, 768 (2002).

\bibitem{Shreiber03}
D.~Shreiber, V.~Barocas, and R.~Tranquillo,
\newblock Biophys. J. {\bf 84}, 4102 (2003).

\bibitem{Sambeth01a}
R.~Sambeth and A.~Baumgaertner,
\newblock Phys. Rev. Lett. {\bf 86}, 5196 (2001).

\bibitem{Carlier00}
D.~Pantaloni, R.~Boujemaa, D.~Didry, P.~Gounon, and M.-F. Carlier,
\newblock Nature Cell Biology {\bf 2}, 385 (2000).

\bibitem{Vicker00}
M.~G. Vicker,
\newblock Biophysical Chemistry {\bf 84}, 87 (2000).

\end{thebibliography}

\end{document}